\documentclass[aps,prl,twocolumn,showpacs]{revtex4}
\usepackage{graphicx}

\begin{document}
\title{Symmetries of the Resistance of Mesoscopic Samples in the Quantum Hall Regime 
}
\author{E. Peled$^{1}$}
\author{Y. Chen$^{2}$}
\author{E. Diez$^{2}$}
\thanks{Present address: Departamento de F\'{i}sica, Facultad de Ciencias, Universidad de Salamanca, E-37008 Salamanca, Spain}
\author{D. C. Tsui$^{2}$}
\author{D. Shahar$^{1}$}
\author{D. L. Sivco$^{3}$}
\author{A. Y. Cho$^{3}$}
\affiliation{
$^{1}$Department of Condensed Matter Physics, The Weizmann Institute of Science, Rehovot 
76100, Israel,\\
$^{2}$Department of Electrical Engineering, Princeton University, 
Princeton, New Jersey 08544, USA,\\
$^{3}$Bell Laboratories, Lucent Technologies, 600 Mountain Avenue, Murray Hill, 
New Jersey 07974, USA.
}

\begin{abstract}
The symmetry properties of the resistance of mesoscopic samples in the quantum Hall regime are investigated. In addition to the reciprocity relation, our samples obey new symmetries, that relate resistances measured with different contact configurations. Different kinds of symmetries are identified, depending on whether the magnetic field value is such that the system is above, or below, a quantum Hall transition. Related symmetries have recently been reported for macroscopic samples in the quantum Hall regime by Ponomarenko {\it et al.} (Solid State Commun.\ {\bf 130}, 705 (2004)), and Karmakar {\it et al.} (Preprint cond-mat/0309694).
\end{abstract}


\maketitle


Early studies of mesoscopic conductance fluctuations in the presence of a magnetic field ($B$) revealed an apparent puzzle \cite{Umbach1984PRB30,Webb1985PRL54_2}: the pattern of fluctuations obtained from thin-film metallic samples exhibited no specific symmetry with respect to the reversal of $B$. 
These findings appeared puzzling because the conductivity of the samples was expected to follow the Onsager relations \cite{Onsager1931PR38,Casimir1945RMP17}, $\sigma_{\alpha \beta}(B)=\sigma_{\beta \alpha}(-B)$, where $\alpha$ and $\beta$ refer to coordinates, and therefore to have a clear symmetry upon $B$ reversal. 
This apparent contradiction was soon settled by Benoit {\it et al.} \cite{Benoit1986PRL57}, and B\"{u}ttiker \cite{Buttiker1986PRL57,Buttiker1988IBM32},  who derived a general formula for the experimentally measured four-terminal resistance configuration and demonstrated that this resistance (or conductance) need not be symmetric with respect to the reversal of $B$. Instead, it should obey the reciprocity relation, stating that it will be symmetric with respect to the reversal of $B$ and the simultaneous exchange of the current and voltage contacts: 
\begin{equation}
    R_{ij,kl}(B)=R_{kl,ij}(-B). 
\end{equation}
Here we use the standard notation of $R_{ij,kl}$ for $V_{kl}/I_{ij}$, where $V_{kl}$ is the voltage difference between contacts $k$ and $l$ and $I_{ij}$ is the current between contacts $i$ and $j$. 

These experiments, performed using metallic wires and loops, were limited to the low-$B$ regime where Landau levels are unresolved. We present, in this Communication, an experimental study of the resistance of mesoscopic samples, designed to test their symmetries in the quantum Hall (QH) regime. We show that, in addition to the reciprocity relation, resistances measured near transitions between QH states exhibit symmetries that are not predicted by B\"{u}ttiker's four-terminal resistance formula. These symmetries describe relations between resistances that are obtained with different contact configurations.

\begin{figure}[ht]
\includegraphics[scale=0.48]{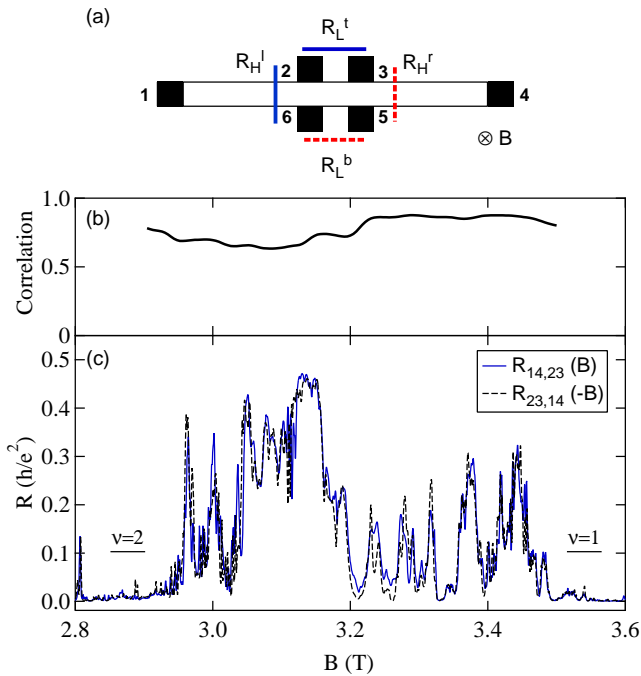} 
\caption{(Color online) 
(a) Geometry and contact numbering of our samples. The black areas in the figure represent Au-Ge-Ni alloyed contacts that were designed to reach the edges of the Hall-bar. The Hall-bar has a lithographic width of 2 $\mu$m, with a center-to-center distance of 4 $\mu$m between the longitudinal voltage contacts (2 to 3 and 6 to 5) and a distance of 24 $\mu$m between the current contacts (1 to 4). The four voltage contact-pairs that are used in the measurements are denoted by the corresponding resistances, $R_{L}^{t}$, $R_{L}^{b}$, $R_{H}^{l}$, and $R_{H}^{r}$.
(b) The correlation between the fluctuations of $R_{14,23}(B)$ and $R_{23,14}(-B)$.
(c) The longitudinal resistance, $R_{14,23}(B)$ ($R_{L}^{t}$ of Fig.\ 1(a)), of sample T2Cm2 at the vicinity of the $\nu=2-1$ transition, together with its reciprocity-equivalent resistance, $R_{23,14}(-B)$ (not shown in Fig.\ 1(a)).   
}
\label{fig:Fig1}
\end{figure}
Our samples were prepared from two InGaAs/InAlAs wafers that contain a 200 \AA \ quantum-well. A two-dimensional electronic system is formed in the quantum-well after illumination with an LED. Due to the short-range alloy scattering in our material the electronic system has a low mobility, limiting our study to the integer QH effect. The data were obtained from two samples, T2C and T1B. Sample T2C was cooled twice (T2Cm2 and T2Cn2), and had a density and mobility of $n_{s}=1.15 \cdot 10^{11}$ cm$^{-2}$, $\mu=14,000$ cm$^{2}$/Vsec for both cool-downs. Sample T1B was cooled once (T1Bc2), and had $n_{s}=3.65 \cdot 10^{11}$ cm$^{-2}$, $\mu=44,000$ cm$^{2}$/Vsec. The samples were wet-etched to a Hall-bar geometry shown in Fig.\ 1(a). Special care was taken in the alignment of the metallic contacts to the Hall-bars, to ensure that voltage probes on opposite sides of a Hall-bar will probe the same region of the sample.
Due to the small size of our samples, their resistances display reproducible fluctuations whose magnitude and $B$-correlations near $B=0$ were used to extract the phase-coherence length, $L_{\phi}$ \cite{Lee1987PRB35}. For our samples $L_{\phi}=1.1-1.3\ \mu$m at a temperature ($T$) of 10 mK, the $T$ at which all of the data presented here were taken.
Four-terminal resistance measurements were done using standard ac lock-in techniques with frequencies of 3 -- 4 Hz and a current $I=1$ nA, safely below $I=10$ nA where the fluctuations begin to diminish in size. $B$-field sweep-rates were from 0.02 to 0.05 T/min, keeping the fluctuations independent of sweep rate.

We begin the presentation of our data with an experimental test of the reciprocity relation for our samples in the QH regime. In Fig.\ 1(c) we compare two reciprocity-equivalent resistances of sample T2Cm2 in the vicinity of the transition from the $\nu=2$ to the  $\nu=1$ QH state ($\nu=2-1$ transition, where $\nu$ is the Landau level filling factor). Referring to the contact numbering in Fig.\ 1(a), the resistances compared in Fig.\ 1(c) are a longitudinal resistance, $R_{14,23}$, measured at positive $B$, and the resistance obtained after exchanging the current and voltage contacts, $R_{23,14}$, and taken at negative $B$ polarity. Although the resistances are dominated by reproducible fluctuations we find that, in accordance with the reciprocity relation of Eq.\ 1, they have nearly the same pattern differing by only a small fraction of their amplitude.
In order to quantify their similarity, we calculate the correlation between the fluctuations of the two resistances, normalized by the autocorrelation of each fluctuation pattern \cite{Overlap,Skocpol1987PRL58}, and averaged over a $B$ range of 0.2 T. In Fig.\ 1(b) we plot the results of the correlation calculation for the data of Fig.\ 1(c). The high values of the correlation, ranging from 0.63 to 0.88, attest to the similarity between $R_{14,23}(B)$ and $R_{23,14}(-B)$.
Other resistances that are related via Eq.\ 1, such as $R_{14,62}(B)$ and $R_{62,14}(-B)$, were also found to have similarly high correlation-values. While the reciprocity relation has been demonstrated before for the QH regime \cite{Sample1987JAP61}, we are extending it here to samples whose resistances are dominated by mesoscopic fluctuations.

\begin{figure}[ht]
\includegraphics[scale=0.48]{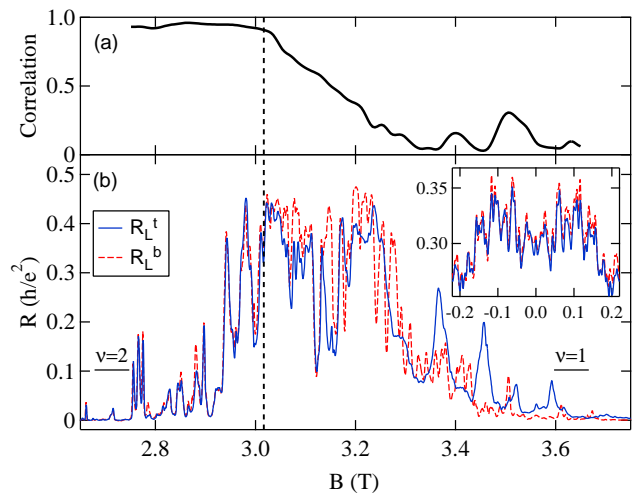} 
\caption{(Color online) 
(a) The correlation between the fluctuations of $R_{L}^{t}$ and $R_{L}^{b}$. \\
(b) $R_{L}^{t}$ and $R_{L}^{b}$ vs.\ $B$ obtained from sample T2Cn2 in the vicinity of the $\nu=2-1$ transition. Inset: The same resistances near $B=0$.
}
\label{fig:Fig2}
\end{figure}
The main purpose of this Communication is to describe new symmetry properties, of four-terminal measurements, that are particular to the QH regime. These symmetries relate resistance measurements done with different contact configurations using samples in the mesoscopic regime. For simplicity we compare the two Hall and two longitudinal measurement-configurations illustrated in Fig.\ 1(a). The current flows between contacts 1 and 4, and voltages are measured using the contact pairs 2-3 for the `top' longitudinal resistance ($R_{L}^{t}$) and 6-5 for the `bottom' longitudinal resistance ($R_{L}^{b}$). Similarly, we use the contact pairs 6-2 for the `left' and 5-3 for the `right' Hall resistances, $R_{H}^{l}$ and $R_{H}^{r}$, respectively. 
For an ideal, macroscopic and homogeneous, sample there should be no difference between the `top' and `bottom', or `left' and `right', measurements and one would expect to find $R_{L}^{t}=R_{L}^{b}$ and $R_{H}^{l}=R_{H}^{r}$. 
In experiments this is rarely the case and, in general, each contact configuration yields a different result. In the following we show that, although our measurements can yield very different results depending on the specific configuration, they are still linked by clear symmetry relations. These relations can involve the simultaneous exchange of contacts and $B$ polarity, in a manner which is akin to that prescribed by the reciprocity relation.

Let us begin by considering the longitudinal configurations. In Fig.\ 2(b) we compare the $R_{L}^{t}$ and $R_{L}^{b}$ obtained from sample T2Cn2, near the $\nu=2-1$ QH transition. The style of the curves (solid or dashed line) corresponds to the voltage-contacts used, as indicated in Fig.\ 1(a). The first observation we make from these data is that they can be divided into two $B$-ranges according to the similarity between $R_{L}^{t}$ and $R_{L}^{b}$. For $B<3.016$ T, on the left side of the dashed line in the figure, $R_{L}^{t}$ and $R_{L}^{b}$ are virtually indistinguishable and their correlation, shown in Fig.\ 2(a), is close to unity. This is similar to the behavior we observe outside the QH regime, for the resistance fluctuations near $B=0$, see the inset of Fig.\ 2(b). 

\begin{figure}[ht]
\includegraphics[scale=0.48]{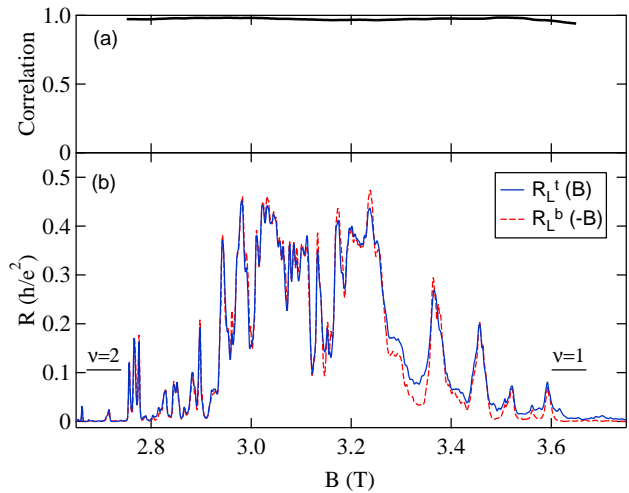} 
\caption{(Color online) (a) The correlation between the fluctuations of $R_{L}^{t}(B)$ and $R_{L}^{b}(-B)$. \\
(b) $R_{L}^{t}(B)$ and $R_{L}^{b}(-B)$ of sample T2Cn2 in the vicinity of the $\nu=2-1$ transition.
}
\label{fig:Fig3}
\end{figure}
The picture changes dramatically as $B$ is increased beyond 3.016 T. The $R_{L}$ traces gradually begin to deviate from each other and eventually become uncorrelated. This can be clearly seen in Fig.\ 2(a), where we plot the correlation function between the two $R_{L}$'s. On the low-$B$ side of the transition, for $B<3.016$ T, the correlation is between 0.89 to 0.96, while on the high-$B$ side, approximately above 3.3 T, it randomly fluctuates between 0.03 to 0.3, indicating the uncorrelated nature of the $R_{L}$'s. At the intermediate $B$ range, 3.016--3.3 T, the correlation interpolates between these two regions. 

The surprising result of our work is that, despite the uncorrelated appearance of the two $R_{L}$ traces at the high-$B$ side of the transition, they do not represent independent measurements. Instead we find that, upon the reversal of $B$, the `top' and `bottom' measurements are mapped onto each other. This is shown in Fig.\ 3(b) where we plot $R_{L}^{t}(B)$ together with $R_{L}^{b}(-B)$. The similarity of the traces is clearly improved and the correlation (Fig.\ 3(a)) is close to unity for the entire range of $B$. We thus identified a new symmetry for mesoscopic samples in the QH regime: 
\begin{equation}
\label{newSymm}
R_{L}^{t}(B)=R_{L}^{b}(-B).
\end{equation}
This symmetry holds also for the low-$B$ side of the transition, indicating that $R_{L}$ in this region is symmetric in $B$. 

\begin{figure}[ht]
\includegraphics[scale=0.48]{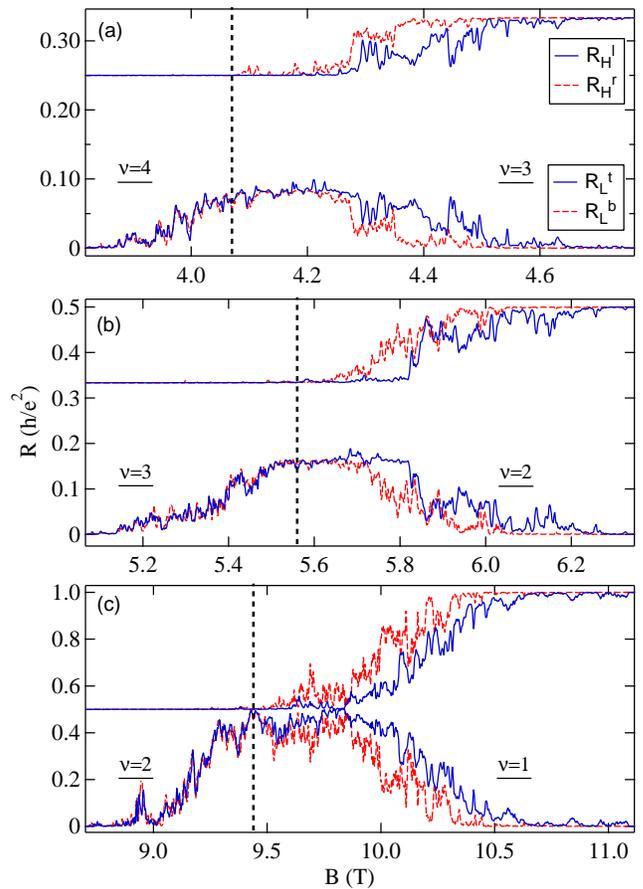} 
\caption{(Color online) 
$R_{L}$ (bottom traces in each graph) and $R_{H}$ (top traces) vs.\ $B$ obtained from sample T1Bc2 in the vicinity of the $\nu=4-3$ (a), $3-2$ (b), and $2-1$ (c) QH transitions. 
}
\label{fig:Fig4}
\end{figure}
To check whether this behavior is common to other QH transitions we repeated our measurements with a higher density sample, T1Bc2, allowing us to observe well-separated $\nu=4-3$, $3-2$, and $2-1$ QH transitions. 
The bottom pair of traces in each graph in Fig.\ 4 are measurements of $R_{L}^{t}$ and $R_{L}^{b}$ near each transition. The division into high- and low-$B$ ranges is evident for all transitions studied. We have also verified the validity of Eq.\ \ref{newSymm} for these transitions.

The symmetry of Eq.\ \ref{newSymm} has been observed before in large, macroscopic, Hall-bar shaped samples, and has been attributed to the existence of a longitudinal density-gradient in the samples \cite {Klitzing1984TDS}.
Recently, Ponomarenko {\it et al.} and Karmakar {\it et al.} \cite{PonomarenkoCM} reported on the observation of this symmetry and suggested a model that details how the symmetry originates from a density gradient. Due to a longitudinal density-gradient in their samples the `left' and `right' Hall resistances of the samples are not equal, $R_{H}^{l}\neq R_{H}^{r}$, and are instead found to be $B$-shifted with respect to each other. This then leads to the existence of a difference between the longitudinal resistances, $R_{L}^{t}$ and $R_{L}^{b}$, since according to Kirchhoff's law the Hall voltage difference is equivalent to a longitudinal voltage difference, or: $R_{H}^{r}-R_{H}^{l}=R_{L}^{t}-R_{L}^{b}$. In their model, the authors of Ref.\ \cite{PonomarenkoCM} calculate the various $R_{L}$'s and $R_{H}$'s of the sample, with their explicit dependence on the density gradient, and show that for a linear density gradient the $R_{L}$'s follow the symmetry of Eq.\ \ref{newSymm}.

In the following paragraph we discuss the properties of $R_{H}$ in our samples. We show that in our samples the differences between $R_{H}^{l}$ and $R_{H}^{r}$ do not amount to only a $B$-shift between the two measured traces. Each one of the $R_{H}$ measurements displays a distinctly different pattern of fluctuations, that are related to the $R_{L}$ fluctuations. The appearance of such fluctuation-dominated resistances that, nonetheless, obey the symmetry of Eq.\ \ref{newSymm} can not be accounted for by a density gradient, indicating that our observations are not within the scope of the model suggested in \cite{PonomarenkoCM}.

In Fig.\ 4 we present, along with the $R_{L}$ traces, the $R_{H}$ measurements corresponding to each transition. First, we note that whenever $R_{L}^{t}=R_{L}^{b}$, as in the case for the low-$B$ side of the data in the figure, $R_{H}^{l}=R_{H}^{r}$ must follow. Inspecting the $R_{H}$'s in the figure we see that not only they are equal but they are also in their quantized state, $R_{H}^{l}=R_{H}^{r}=h/ie^{2}$, where $i$ is 2, 3 or 4. As $B$ is increased through the transition, and the $R_{L}$'s begin to separate, the $R_{H}$'s, maintaining Kirchhoff's law, separate as well and cease to be quantized. At this higher-$B$ range, described in detail in a previous publication \cite{Peled2003PRL91}, we have found another kind of correlation, between the fluctuations of $R_{L}$ and those of $R_{H}$:
\begin{equation}
    R_{L}^{t}+R_{H}^{l}=R_{L}^{b}+R_{H}^{r}=h/(i-1)e^{2}.
\end{equation}
We emphasize that these correlations appear only between specific $R_{L}$-$R_{H}$ pairs, depending on the polarity of $B$: At positive $B$ the correlated pairs are $R_{L}^{t}$-$R_{H}^{l}$, and $R_{L}^{b}$-$R_{H}^{r}$ (the styles of the traces in Fig.\ 4 were chosen to highlight these correlations), while at negative $B$ they are $R_{L}^{t}$-$R_{H}^{r}$ and $R_{L}^{b}$-$R_{H}^{l}$. The switching of pairs at negative $B$ is a result of our samples having $R_{H}$'s that are antisymmetric with respect to the reversal of $B$, together with the $B$-symmetry of $R_{L}$, Eq.\ \ref{newSymm}.

Our findings can be summarized as follows: for a transition from a $\nu = i$ to a $\nu = i-1$ QH state the low-$B$ side of the transition has $R_{L}^{t}=R_{L}^{b}$ and $R_{H}^{l}=R_{H}^{r}$. $R_{L}^{t}$ and $R_{L}^{b}$ are non-zero and exhibit fluctuations, while $R_{H}^{l}$ and $R_{H}^{r}$ are quantized to the value of the preceding QH plateau, $h/ie^{2}$.
On the high-$B$ side of the transition $R_{L}^{t}\neq R_{L}^{b}$ and $R_{H}^{l}\neq R_{H}^{r}$. $R_{L}^{t (b)}$ and $R_{H}^{l (r)}$ are anti-correlated, exhibiting fluctuations of equal magnitude and opposite sign. Their sum equals the resistance value of the next QH plateau, $h/(i-1)e^{2}$. 
The $B$-symmetry of the resistances can be neatly summarized by the following observation: For both $R_{L}$ and $R_{H}$ the effect of reversing the direction of $B$ is equivalent to an exchange of the `top' and `bottom' voltage probes.

In our samples the QH series is terminated with a transition to an insulating phase as $B$ is increased beyond the $\nu=1$ QH state. In the vicinity of this last transition, on which we have reported previously \cite{Peled2003PRL90}, the behavior of $R_{L}$ and $R_{H}$ is similar to that observed for the low-$B$ side of the higher-LL transitions. Both $R_{L}$'s are dominated by reproducible fluctuations that are nearly equal, $R_{L}^{t}=R_{L}^{b}$, and the Hall resistances are quantized to $h/e^{2}$, their value at the QH state preceding the transition, $R_{H}^{l}=R_{H}^{r}=h/e^{2}$.

The accepted theoretical model for describing transport in mesoscopic samples at the QH regime is based on B\"{u}ttiker's four-terminal resistance formula extended to include the existence of electronic edge-states \cite{Buttiker1988PRB38}. The presence of edge-states, whose chirality is determined by the polarity of $B$, may point in the direction of the origin of the symmetries presented in this Communication.
However, our observed symmetries in the QH regime do not emerge from a straightforward application of the B\"{u}ttiker multi-probe formula. 

In the transport models of Streda, Kucera and MacDonald \cite{Streda1987PRL59} and of Jain and Kivelson \cite{Jain1988PRL60} resistance fluctuations appear as a result of electrons scattering between the `top' and `bottom' edge states. When only one edge state is present, corresponding to conduction via the lowest Landau level alone, these models predict that the fluctuations will be limited to $R_{L}$, leaving $R_{H}$ quantized. This situation is in agreement with our observations at the transition to the insulating phase, but does not account for the quantized $R_{H}$ on the low-$B$ side of the higher Landau level transitions of $\nu=4-3$, $3-2$, and $2-1$.

In a recent numerical simulation \cite{Zhou2003CM0311365} Zhou and Berciu  make use of B\"{u}ttiker's formulation to describe the resistance in the QH regime as a result of an interplay between chiral edge-currents and the tunneling between the `top' and `bottom' edges of the Hall-bar. Their simulations reproduce many of the central features of our results, identifying a low- and high-$B$ regions for all QH transitions, and predicting the symmetry of Eq.\ \ref{newSymm}.  According to their model, on the low-$B$ side of the transitions transport is dominated by the presence of edge-states, together with tunneling between the `top' and `bottom' sides of the sample, while on the high-$B$ side transport is enabled only via the latter process, with no edge states or tunneling between the `left' and `right' sides of the sample.

To conclude, we presented an experimental study of the symmetries of the resistance of mesoscopic samples in the QH regime. We demonstrated new symmetries, relating longitudinal and Hall resistances of different contact configurations and $B$ polarities. The resistances in the vicinity of all QH transitions were found to follow one of two possible sets of symmetries, one on the low-$B$ and the other on the high-$B$ side of the transitions.

We wish to thank useful discussions with M.\ B\"{u}ttiker, Y.\ Oreg, A.\ Stern, and C.\ Zhou. This work is supported by the BSF and by the Koshland Fund. Y.\ C.\ is supported by the (U.S.) NSF. E.\ D.\ is supported by the Ram\'{o}n y Cajal Program of the Spanish Minister of Science and Technology.


\begin{thebibliography}{19}
\expandafter\ifx\csname natexlab\endcsname\relax\def\natexlab#1{#1}\fi
\expandafter\ifx\csname bibnamefont\endcsname\relax
  \def\bibnamefont#1{#1}\fi
\expandafter\ifx\csname bibfnamefont\endcsname\relax
  \def\bibfnamefont#1{#1}\fi
\expandafter\ifx\csname citenamefont\endcsname\relax
  \def\citenamefont#1{#1}\fi
\expandafter\ifx\csname url\endcsname\relax
  \def\url#1{\texttt{#1}}\fi
\expandafter\ifx\csname urlprefix\endcsname\relax\def\urlprefix{URL }\fi
\providecommand{\bibinfo}[2]{#2}
\providecommand{\eprint}[2][]{\url{#2}}

\bibitem[{\citenamefont{Umbach et~al.}(1984)\citenamefont{Umbach, Washburn,
  Laibowitz, and Webb}}]{Umbach1984PRB30}
\bibinfo{author}{\bibfnamefont{C.~P.} \bibnamefont{Umbach}},
  \bibinfo{author}{\bibfnamefont{S.}~\bibnamefont{Washburn}},
  \bibinfo{author}{\bibfnamefont{R.~B.} \bibnamefont{Laibowitz}},
  \bibnamefont{and} \bibinfo{author}{\bibfnamefont{R.~A.} \bibnamefont{Webb}},
  \bibinfo{journal}{Phys. Rev. B} \textbf{\bibinfo{volume}{30}},
  \bibinfo{pages}{4048} (\bibinfo{year}{1984}).

\bibitem[{\citenamefont{Webb et~al.}(1985)\citenamefont{Webb, Washburn, Umbach,
  and Laibowitz}}]{Webb1985PRL54_2}
\bibinfo{author}{\bibfnamefont{R.~A.} \bibnamefont{Webb}},
  \bibinfo{author}{\bibfnamefont{S.}~\bibnamefont{Washburn}},
  \bibinfo{author}{\bibfnamefont{C.~P.} \bibnamefont{Umbach}},
  \bibnamefont{and} \bibinfo{author}{\bibfnamefont{R.~B.}
  \bibnamefont{Laibowitz}}, \bibinfo{journal}{Phys. Rev. Lett.}
  \textbf{\bibinfo{volume}{54}}, \bibinfo{pages}{2696} (\bibinfo{year}{1985}).

\bibitem[{\citenamefont{Onsager}(1931)}]{Onsager1931PR38}
\bibinfo{author}{\bibfnamefont{L.}~\bibnamefont{Onsager}},
  \bibinfo{journal}{Phys. Rev.} \textbf{\bibinfo{volume}{38}},
  \bibinfo{pages}{2265} (\bibinfo{year}{1931}).

\bibitem[{\citenamefont{Casimir}(1945)}]{Casimir1945RMP17}
\bibinfo{author}{\bibfnamefont{H.~B.~G.} \bibnamefont{Casimir}},
  \bibinfo{journal}{Rev. Mod. Phys.} \textbf{\bibinfo{volume}{17}},
  \bibinfo{pages}{343} (\bibinfo{year}{1945}).

\bibitem[{\citenamefont{Benoit et~al.}(1986)\citenamefont{Benoit, Washburn,
  Umbach, Laibowitz, and Webb}}]{Benoit1986PRL57}
\bibinfo{author}{\bibfnamefont{A.~D.} \bibnamefont{Benoit}},
  \bibinfo{author}{\bibfnamefont{S.}~\bibnamefont{Washburn}},
  \bibinfo{author}{\bibfnamefont{C.~P.} \bibnamefont{Umbach}},
  \bibinfo{author}{\bibfnamefont{R.~B.} \bibnamefont{Laibowitz}},
  \bibnamefont{and} \bibinfo{author}{\bibfnamefont{R.~A.} \bibnamefont{Webb}},
  \bibinfo{journal}{Phys. Rev. Lett.} \textbf{\bibinfo{volume}{57}},
  \bibinfo{pages}{1765} (\bibinfo{year}{1986}).

\bibitem[{\citenamefont{B\"{u}ttiker}(1986)}]{Buttiker1986PRL57}
\bibinfo{author}{\bibfnamefont{M.}~\bibnamefont{B\"{u}ttiker}},
  \bibinfo{journal}{Phys. Rev. Lett.} \textbf{\bibinfo{volume}{57}},
  \bibinfo{pages}{1761} (\bibinfo{year}{1986}).

\bibitem[{\citenamefont{B\"{u}ttiker}(1988{\natexlab{a}})}]{Buttiker1988IBM32}
\bibinfo{author}{\bibfnamefont{M.}~\bibnamefont{B\"{u}ttiker}},
  \bibinfo{journal}{IBM J. Res. Dev.} \textbf{\bibinfo{volume}{32}},
  \bibinfo{pages}{317} (\bibinfo{year}{1988}{\natexlab{a}}).

\bibitem[{\citenamefont{Lee et~al.}(1987)\citenamefont{Lee, Stone, and
  Fukuyama}}]{Lee1987PRB35}
\bibinfo{author}{\bibfnamefont{P.~A.} \bibnamefont{Lee}},
  \bibinfo{author}{\bibfnamefont{A.~D.} \bibnamefont{Stone}}, \bibnamefont{and}
  \bibinfo{author}{\bibfnamefont{H.}~\bibnamefont{Fukuyama}},
  \bibinfo{journal}{Phys. Rev. B} \textbf{\bibinfo{volume}{35}},
  \bibinfo{pages}{1039} (\bibinfo{year}{1987}).

\bibitem[{Ove()}]{Overlap}
\bibinfo{note}{We define the normalized correlation between the fluctuations of
  any two resistance measurements, $R^{(1)}$ and $R^{(2)}$, as: $correlation
  =|cor(\delta R^{(1)}, \delta R^{(2)})/\sqrt{cor(\delta R^{(1)}, \delta
  R^{(1)})\times cor(\delta R^{(2)}, \delta R^{(2)})}|$. Here $cor$ stands for
  the value of the cross-correlation function at zero shift.}

\bibitem[{\citenamefont{Skocpol et~al.}(1987)\citenamefont{Skocpol, Mankiewich,
  Howard, Jackel, Tennant, and Stone}}]{Skocpol1987PRL58}
\bibinfo{author}{\bibfnamefont{W.~J.} \bibnamefont{Skocpol}},
  \bibinfo{author}{\bibfnamefont{P.~M.} \bibnamefont{Mankiewich}},
  \bibinfo{author}{\bibfnamefont{R.~E.} \bibnamefont{Howard}},
  \bibinfo{author}{\bibfnamefont{L.~D.} \bibnamefont{Jackel}},
  \bibinfo{author}{\bibfnamefont{D.~M.} \bibnamefont{Tennant}},
  \bibnamefont{and} \bibinfo{author}{\bibfnamefont{A.~D.} \bibnamefont{Stone}},
  \bibinfo{journal}{Phys. Rev. Lett.} \textbf{\bibinfo{volume}{58}},
  \bibinfo{pages}{2347} (\bibinfo{year}{1987}).

\bibitem[{\citenamefont{Sample et~al.}(1987)\citenamefont{Sample, Bruno,
  Sample, and Sichel}}]{Sample1987JAP61}
\bibinfo{author}{\bibfnamefont{H.~H.} \bibnamefont{Sample}},
  \bibinfo{author}{\bibfnamefont{W.~J.} \bibnamefont{Bruno}},
  \bibinfo{author}{\bibfnamefont{S.~B.} \bibnamefont{Sample}},
  \bibnamefont{and} \bibinfo{author}{\bibfnamefont{E.~K.}
  \bibnamefont{Sichel}}, \bibinfo{journal}{J. Appl. Phys.}
  \textbf{\bibinfo{volume}{61}}, \bibinfo{pages}{1079} (\bibinfo{year}{1987}).

\bibitem[{Kli()}]{Klitzing1984TDS}
\bibinfo{note}{K. v. Klitzing and G. Ebert, in {\it Two-Dimensional Systems,
  Heterostructures, and Superlattices}, Springer Series in Solid-State
  Sciences, edited by G. Bauer, F. Kuchar, and H. Heinrich, (Springer-Verlag,
  Berlin 1984), Vol. 53, p. 242}.

\bibitem[{Pon()}]{PonomarenkoCM}
\bibinfo{note}{L. A. Ponomarenko, D. T. N. de Lang, A. de Visser, V. A.
  Kulbachinskii, G. B. Galiev, H. K\"{u}nzel, and A. M. M. Pruisken, Solid
  State Commun. {\bf 130}, 705 (2004); B. Karmakar, M. R. Gokhale, A. P. Shah,
  B. M. Arora, D. T. N. de Lang, A. de Visser, L. A. Ponomarenko, and A. M. M.
  Pruisken, Preprint cond-mat/0309694.}

\bibitem[{\citenamefont{Peled et~al.}(2003{\natexlab{a}})\citenamefont{Peled,
  Shahar, Chen, Diez, Sivco, and Cho}}]{Peled2003PRL91}
\bibinfo{author}{\bibfnamefont{E.}~\bibnamefont{Peled}},
  \bibinfo{author}{\bibfnamefont{D.}~\bibnamefont{Shahar}},
  \bibinfo{author}{\bibfnamefont{Y.}~\bibnamefont{Chen}},
  \bibinfo{author}{\bibfnamefont{E.}~\bibnamefont{Diez}},
  \bibinfo{author}{\bibfnamefont{D.~L.} \bibnamefont{Sivco}}, \bibnamefont{and}
  \bibinfo{author}{\bibfnamefont{A.~Y.} \bibnamefont{Cho}},
  \bibinfo{journal}{Phys. Rev. Lett.} \textbf{\bibinfo{volume}{91}},
  \bibinfo{pages}{236802} (\bibinfo{year}{2003}{\natexlab{a}}).

\bibitem[{\citenamefont{Peled et~al.}(2003{\natexlab{b}})\citenamefont{Peled,
  Shahar, Chen, Sivco, and Cho}}]{Peled2003PRL90}
\bibinfo{author}{\bibfnamefont{E.}~\bibnamefont{Peled}},
  \bibinfo{author}{\bibfnamefont{D.}~\bibnamefont{Shahar}},
  \bibinfo{author}{\bibfnamefont{Y.}~\bibnamefont{Chen}},
  \bibinfo{author}{\bibfnamefont{D.~L.} \bibnamefont{Sivco}}, \bibnamefont{and}
  \bibinfo{author}{\bibfnamefont{A.~Y.} \bibnamefont{Cho}},
  \bibinfo{journal}{Phys. Rev. Lett.} \textbf{\bibinfo{volume}{90}},
  \bibinfo{pages}{246802} (\bibinfo{year}{2003}{\natexlab{b}}).

\bibitem[{\citenamefont{B\"{u}ttiker}(1988{\natexlab{b}})}]{Buttiker1988PRB38}
\bibinfo{author}{\bibfnamefont{M.}~\bibnamefont{B\"{u}ttiker}},
  \bibinfo{journal}{Phys. Rev. B} \textbf{\bibinfo{volume}{38}},
  \bibinfo{pages}{9375} (\bibinfo{year}{1988}{\natexlab{b}}).

\bibitem[{\citenamefont{Streda et~al.}(1987)\citenamefont{Streda, Kucera, and
  MacDonald}}]{Streda1987PRL59}
\bibinfo{author}{\bibfnamefont{P.}~\bibnamefont{Streda}},
  \bibinfo{author}{\bibfnamefont{J.}~\bibnamefont{Kucera}}, \bibnamefont{and}
  \bibinfo{author}{\bibfnamefont{A.~H.} \bibnamefont{MacDonald}},
  \bibinfo{journal}{Phys. Rev. Lett.} \textbf{\bibinfo{volume}{59}},
  \bibinfo{pages}{1973} (\bibinfo{year}{1987}).

\bibitem[{\citenamefont{Jain and Kivelson}(1988)}]{Jain1988PRL60}
\bibinfo{author}{\bibfnamefont{J.~K.} \bibnamefont{Jain}} \bibnamefont{and}
  \bibinfo{author}{\bibfnamefont{S.~A.} \bibnamefont{Kivelson}},
  \bibinfo{journal}{Phys. Rev. Lett.} \textbf{\bibinfo{volume}{60}},
  \bibinfo{pages}{1542} (\bibinfo{year}{1988}).

\bibitem[{\citenamefont{Zhou and Berciu}()}]{Zhou2003CM0311365}
\bibinfo{author}{\bibfnamefont{C.}~\bibnamefont{Zhou}} \bibnamefont{and}
  \bibinfo{author}{\bibfnamefont{M.}~\bibnamefont{Berciu}},
  \bibinfo{journal}{Preprint cond-mat/0311365}.

\end{thebibliography}

\end{document}